\documentclass[useAMS,usenatbib,twocolumn]{mn2e}
\usepackage{natbib}
\usepackage{graphicx}
\usepackage{times}
\usepackage{rotate}
\usepackage{color,url}
\newcommand{\prd}{Phys.~Rev.~D}
\newcommand{\aap}{A.\&Ap.}
\newcommand{\mnras}{MNRAS}
\newcommand{\apj}{ApJ}
\newcommand{\aj}{AJ}
\newcommand{\apjl}{ApJ}

\newcommand{\apjs}{ApJS}
\newcommand{\physrep}{Physics Report}

\newcommand{\cgs}{ ${\rm erg~cm}^{-2}~{\rm s}^{-1}$} 

\def\xmm{{XMM-{\it Newton}}}
\def\chandra{{\it Chandra}}

\newcommand{\eg}{{\it e.g.,}}

\begin{document}
  

 \title[Dark Matter Halos of X-ray AGN]{The Dark Matter Halos of
   Moderate Luminosity X-ray AGN as Determined from Weak Gravitational
   Lensing and Host Stellar Masses}




 \author[Leauthaud et al.]
{Alexie Leauthaud$^1$, Andrew J, Benson$^2$, Francesca
  Civano$^3$, Alison L, Coil$^4$, Kevin Bundy$^1$, \newauthor 
Richard Massey$^5$, Malte Schramm$^1$, Andreas Schulze$^1$, Peter
Capak$^6$, Martin Elvis$^7$, \newauthor 
Andrea Kulier$^8$, Jason Rhodes$^{9,10}$\\
$^1$
Kavli Institute for the Physics and Mathematics of the Universe
(Kavli IPMU, WPI), The University of Tokyo, Chiba 277-8582, Japan\\
$^2$Carnegie Observatories, 813 Santa Barbara Street, Pasadena, CA
91101, USA\\
$^3$Yale Center for Astronomy and Astrophysics, 260 Whitney ave, New Haven, CT 06520, USA\\
$^4$Department of Physics, Center for Astrophysics and Space Sciences,
University of California at San Diego, 9500 Gilman Dr., La Jolla, San
Diego, CA 92093\\
$^5$Institute for Computational Cosmology, Durham University, South Road, Durham DH1 3LE, UK\\
$^6$Spitzer Science Center, 314-6 Caltech, Pasadena, CA, 91125\\
$^7$Harvard Smithsonian Center for astrophysics, 60 Garden St., Cambridge, MA 02138, USA\\
$^8$Department of Astrophysical Sciences, Princeton University,Princeton, NJ 08544, USA\\
$^9$Jet Propulsion Laboratory, California Institute of Technology,
Pasadena, CA 91109\\
$^{10}$California Institute of Technology, Pasadena, CA 91125}

\maketitle
\label{firstpage}

  
\begin{abstract} Understanding the relationship between galaxies
  hosting active galactic nuclei (AGN) and the dark matter halos in
  which they reside is key to constraining how black-hole fueling is
  triggered and regulated.  Previous efforts have relied on simple
  halo mass estimates inferred from clustering, weak gravitational
  lensing, or halo occupation distribution modeling.  In practice,
  these approaches remain uncertain because AGN, no matter how they
  are identified, potentially live a wide range of halo masses with an
  occupation function whose general shape and normalization are poorly
  known. In this work, we show that better constraints can be achieved
  through a rigorous comparison of the clustering, lensing, and
  cross-correlation signals of AGN hosts to the fiducial
  stellar-to-halo mass relation (SHMR) derived for all galaxies,
  irrespective of nuclear activity.  Our technique exploits the fact
  that the global SHMR can be measured with much higher accuracy than
  any statistic derived from AGN samples alone. Using 382 moderate
  luminosity X-ray AGN at $z<1$ from the COSMOS field, we report the
  first measurements of weak gravitational lensing from an X-ray
  selected sample.  Comparing this signal to predictions from the
  global SHMR, we find that, contrary to previous results, most X-ray
  AGN do not live in medium size groups ---nearly half reside in
  relatively low mass halos with $M_{\rm 200b}\sim10^{12.5}$
  M$_\odot$. The AGN occupation function is well described by the same
  form derived for all galaxies but with a lower normalization---the
  fraction of halos with AGN in our sample is a few percent. The
  number of AGN satellite galaxies scales as a power law with host
  halo mass with a power-law index $\alpha=1$. By highlighting the
  relatively ``normal'' way in which moderate luminosity X-ray AGN
  hosts occupy halos, our results suggest that the environmental
  signature of distinct fueling modes for luminous QSOs compared to
  moderate luminosity X-ray AGN is less obvious than previously
  claimed.\end{abstract}

 

 
 


\section{Introduction}

Strong observational evidence suggests a tight coupling between the
growth of super-massive black holes (BHs) and the build-up of galaxy
bulges \citep[][]{Gebhardt:2000,Ferrarese:2000}. In contrast, we only
have a limited understanding of how BH activity relates to dark matter
{\em halo mass} because both halo masses and black hole masses are
challenging to probe observationally. Improved measurements of the BH-dark matter relation
are, however, of great theoretical interest and are key in order to
facilitate a more direct comparison between observations and
theoretical models of Active-Galactic Nuclei (AGN) activity
\citep[\eg][]{DeGraf:2012,Chatterjee:2012,Fanidakis:2013,Hutsi:2014}.

The AGN - halo mass relation is typically probed by measuring the
clustering
\citep[\eg][]{Li:2006,Coil:2009,Gilli:2009,Allevato:2011,Miyaji:2011,Krumpe:2012,
  Koutoulidis:2013,Mountrichas:2013,Shen:2013} or the weak
gravitational lensing of AGN host galaxies
\citep[][]{Mandelbaum:2009}. Halo masses (hereafter $M_{\rm h}$) are
typically inferred from these types of approaches by measuring the
mean large scale bias of a given sample.  Bias values are then
translated into an effective halo mass via the halo mass-bias relation
\citep[\eg][]{Tinker:2010}. However, there are several important
caveats to this approach. First, large scale bias is not a sensitive
probe of halo mass at lower mass scales (the halo mass-bias relation
flattens). Second, converting large scale bias to halo mass requires
assumptions about satellite fractions. Third, the effective halo mass
measured in this fashion corresponds to a bias-weighted average of the
underlying halo mass distribution. For samples which span a wide {\em
  range} of halo masses, there is no simple way to relate this
effective halo mass to more useful averages such as the mode, mean, or
median value of the halo mass distribution.

In principle, these issues can be resolved by adopting an Halo
Occupation Distribution (HOD) type approach which assumes a parametric
model to describe the probability distribution $P(N|M_{\rm h})$ that a
halo of mass $M_{\rm h}$ is host to N galaxies in a given sample
\citep[for a review, see][]{Cooray:2002}. While an HOD type approach
may work well for galaxy samples defined by simple luminosity
thresholds, it is less clear which parametric form should be adopted
for occupation functions when considering AGN-type samples
\citep[\eg][]{Allevato:2011,Miyaji:2011,Kayo:2012,Richardson:2013}. The
AGN duty cycle relative to halos is unknown, which leads to large
uncertainties in both the {\em shape and normalization} of the AGN
occupation functions. Recently, \citet[][]{Shen:2013} measured the
cross-correlation between Quasars (QSOs) and Luminous Red Galaxies
(LRGs) from the Sloan Digital Survey (SDSS) at $z=0.5$. Despite the
high signal-to-noise of their cross-correlation measurement,
\citet[][]{Shen:2013} find that substantially different HODs provide
equally good fits to their data. The conclusions from this work
suggest that clustering data alone is insufficient to fully constrain
the QSO HOD -- underlining the difficulty of modeling AGN type
populations.

For samples of less luminous AGN, such as those selected via deep
X-ray imaging, these issues are even more pronounced. Typical sample
sizes are small, commonly ranging from a few hundred to a few thousand
AGN which means that clustering measurements are noisy. To compensate
for small samples sizes, many studies measure AGN clustering over a
broad range in redshift ($0<z<3$ is not uncommon), X-ray luminosity
(hereafter $L_X$), and host galaxy properties
\citep[\eg][]{Coil:2009,Allevato:2011,Koutoulidis:2013}. Even greater
caution is required when interpreting HODs or bias measurements in
this context.

For moderate luminosity obscured (type-2) AGN samples, however,
information about the properties of the host galaxy contains key
additional information which has yet to be fully exploited for these
types of studies. For obscured systems, the host galaxy light is the
dominant component in the optical/near-infrared Spectral Energy
Distribution (SED), meaning that the stellar mass of the host galaxy
(hereafter $M_*$) can be measured with relatively little contamination
from the AGN component. 

In this paper, we propose an alternative approach to analyzing
clustering and/or lensing measurements of moderate luminosity obscured
AGN samples that can be employed even with small samples. Our approach
relies on using a complete galaxy sample to {\em first} constrain the
overall connection between galaxy mass and halo mass. This model then
serves as a fiducial base-line with which to explore the AGN - halo
mass relation.

From a global perspective that includes all galaxies, tremendous
progress has had been made in recent years in terms of understanding
and modeling the connection between galaxy stellar mass and dark
matter halo mass out to $z=1$ and beyond
\citep[][]{Mandelbaum:2006c,Yang:2009,More:2009,
  Moster:2010,Behroozi:2010,Leauthaud:2011,Leauthaud:2012a}. At the
core of these models is the stellar-to-halo mass (SHMR) for central
galaxies. This may be constrained from measurements of either the
galaxy stellar mass function (SMF), galaxy-clustering, galaxy-galaxy
weak lensing, satellite kinematics, or some combination of these four
probes. In detail, methods vary between different groups, but all
results yield the same global picture: $M_{\rm h}(M_*)$ is well described by
a power-law at low $M_*$ and then transitions to a more sharply rising
function above a characteristic mass scale of $M_*\sim 10^{10.8}$
M$_{\odot}$. The logarithmic scatter in stellar mass at fixed halo
mass is also constrained at $\sigma_{\rm log M_{*}}\sim 0.18$ with
good agreement between different studies. In addition to the SHMR,
these methods also constrain how satellite galaxies populate dark
matter halos as a function of galaxy mass. Finally, the SHMR may also
be constrained as a function of other properties beyond stellar mass,
such as galaxy color or star formation activity
\citep[][]{Mandelbaum:2006c, More:2009, Tinker:2013,Hearin:2013}.

In this paper, we suggest that whenever information about host mass
is available, the AGN-dark matter relation can be probed most
effectively by {\em first} constraining a fiducial SHMR. Once the SHMR
is constrained, the distribution of AGN host stellar masses is all
that is required to make predictions about AGN occupation
statistics. The observed clustering, lensing, or cross-correlations
between AGN and stellar mass limited samples may then be interpreted
in light of predictions from the fiducial SHMR. The key advantage of
this approach is that by using large samples of galaxies that are
complete in terms of stellar mass, the SHMR can be built with much
higher accuracy than by using any statistic measured from AGN samples
alone. Statistics measured from AGN samples (which are necessarily
noisy because of small sample sizes) are only used to constrain {\em
  deviations} from the fiducial model. Any observed deviations would
be of great interest and would provide clues about the mechanisms that
fuel AGN. Our method is similar in many respects to the one adopted by
\citet{Li:2006} and \citet{Mandelbaum:2006c} for analyzing optically
selected and radio-loud AGN.

The approach used here alleviates the difficulties raised by
\citet[][]{Shen:2013} associated with HOD modeling of AGN clustering.
However, it can only be employed for samples with host
stellar mass measurements and therefore cannot be applied in the
context of bright QSO type samples. For these, however, an alternative
and closely related approach has been recently developed by
\citet{Conroy:2013} by combining the SHMR with a BH mass-stellar mass
relation.

We apply our methodology to a sample of X-ray selected moderate
luminosity obscured AGN at $z<1$ from the COSMOS field
\citep[][]{Scoville:2007}. Despite the small sample size (several
hundred AGN) we are able to place robust constraints on AGN halo
occupation statistics. Our choice of the COSMOS field is motivated by
the fact that the galaxy SHMR has been previously constrained for this
field by \citet[][hereafter L12]{Leauthaud:2012a}. The L12 SHMR is
determined from measurements of the galaxy mass function, galaxy
clustering, and galaxy-galaxy lensing to $z=1$. Here, for the first
time, we measure the galaxy-galaxy lensing signal of X-ray selected
obscured AGN. We use this signal to test for differences between the
dark matter environment of obscured AGN compared to the overall galaxy
population.

The layout of this paper is as follows. The data are described in
$\S$\ref{data} followed by the presentation of our methodology in
$\S$\ref{method}. Our main results are presented in
$\S$\ref{results}. Finally, we discuss the results and draw up our
conclusions in $\S$\ref{discussion} and $\S$\ref{conclusions}.

We assume a $\Lambda$CDM cosmology with $\Omega_{\rm m}=0.258$,
$\Omega_\Lambda=0.742$, $\sigma_{8}=0.796$, $H_0=72$
km~s$^{-1}$~Mpc$^{-1}$. All distances are expressed in physical Mpc
units. The letter $M_{\rm h}$ denotes halo mass in general whereas
$M_{\rm 200b}$ is explicitly defined as $M_{\rm 200b}\equiv
M(<r_{\rm 200b})=200\bar{\rho} \frac{4}{3}\pi r_{\rm 200b}^3$ where $r_{\rm 200b}$
is the radius at which the mean interior density is equal to 200 times
the mean matter density ($\bar{\rho}$). Stellar mass is noted $M_{*}$
and has been derived using a Chabrier Initial Mass Function
(IMF). Stellar mass scales as $1/H_0^2$. Halo mass scales as
$1/H_0$. All magnitudes are given on the AB system.


\section{Data and Mock Catalogs}\label{data}

\subsection{COSMOS X-ray AGN Sample}

The AGN sample used for this work is selected by combining the COSMOS
\xmm\ \citep[XMM-COSMOS,][]{Cappelluti:2009} and \chandra\
\citep[C-COSMOS,][]{Elvis:2009} X-ray catalogs. The XMM-COSMOS survey
covers 2 deg$^2$ to a limiting depth of 5$\times$10$^{-16}$ \cgs\ in
the soft (0.5--2 keV) band and 3$\times$10$^{-15}$\cgs\ in the hard
(2--10 keV) band. The C-COSMOS survey covers 0.9 deg$^2$ to a limiting
depth of 1.9$\times$10$^{-16}$ \cgs\ in the soft band and
7.3$\times$10$^{-16}$ \cgs\ in the hard band. The combined catalog of
X-ray sources contains $\sim$1800 objects from XMM-COSMOS and
$\sim$950 objects from C-COSMOS. Details concerning the X-ray
catalogs, the spectroscopic observing programs, and the
spectroscopic/photometric classifications can be found in
\citet{Brusa:2010}, \citet{Civano:2012}, and
\citet{Salvato:2009,Salvato:2011}.

Full band (0.5--10 keV) fluxes are provided in the C-COSMOS catalog
but are not available in the XMM-COSMOS catalog. We compute the full
band flux for XMM-COSMOS sources by summing fluxes in the soft and
hard bands. If a source is not detected in one of the bands, only the
detected flux is included. Rest-frame X-ray luminosities are
homogeneously derived for both catalogs assuming a power law spectral
model with a slope of $\Gamma$=2 and absorption from a Galactic column
density of N$_{\rm H, Gal}$=2.6$\times$10$^{20}$cm$^{-2}$
\citep{Kalberla:2005}. Given that a flat slope has been assumed, no
K-correction is needed.

The aim of this work is to consider moderately obscured and moderate
luminosity AGN for which the host galaxy light is the dominant
component in the optical/near-infrared SED. We select AGN in the
redshift range $0.2<z<1$. All spectroscopically identified broad line
AGN are removed from the sample. A photometric classification
\citep{Salvato:2009,Salvato:2011} is used to identify obscured AGN
when a spectroscopic classification is not available. Spectroscopic
redshifts are available for 71\% of our sample (272/382). We also
impose a lower limit on host mass of $\log_{10}(M_*)>10.5$ (see
section \ref{masses}). This cut is designed to (only very) roughly
match samples used in previous studies of the clustering of X-ray
selected AGN (see section \ref{results}). This mass cut is well above
the COSMOS stellar mass completeness limit at $z=1$ ensuring that our
sample is complete in terms of galaxy mass.

In addition, we also limit our sample to AGN with a rest frame 0.5-10
keV band luminosity in the range $10^{41.5}$ erg s$^{-1}$$<L_X <
10^{43.5}$ erg s$^{-1}$. The upper limit on $L_X $ is set to avoid
bright AGN which might contaminate the host galaxy light. The lower
boundary on $L_X$ is set to limit contamination from star-forming
sources and early type galaxies with pure hot gas X-ray emission
\citep[\eg][]{Civano:2014}. Our results are reasonably robust to
contamination from galaxies outside our sample, provided these span a
similar stellar mass range as our AGN sample. In this case, a 5\%
contamination will have no impact on our mean/median halo mass
estimates, and will simply modify the amplitude of our inferred HOD by
5\%. Our mean/median halo mass estimates are more sensitive to
contamination from galaxies with preferentially low or high $M_*$
values compared to our AGN sample. As an extreme example, if all the
most massive (least massive) galaxies in our sample are contaminants
(at the 5\% level), our mean halo masses will be biased by 8\% (2\%)
and our median halo masses by 7\% (6\%).

\begin{figure}
\begin{center}
\includegraphics[width=8cm]{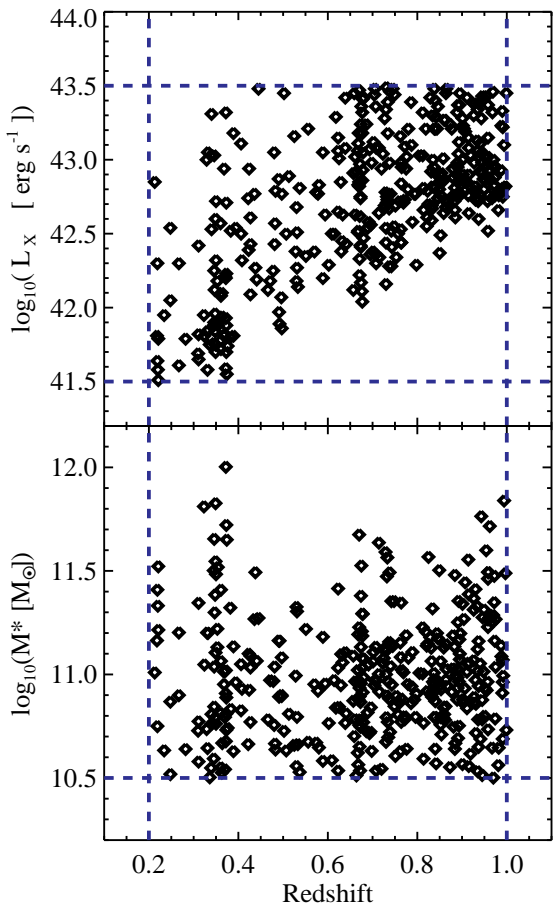}
\caption{Sample of 382 AGN host galaxies as a function of X-ray
  luminosity, stellar mass, and redshift. Our selection cuts are shown
  by the blue dashed lines. The sample is not complete in terms of
  X-ray luminosity but it is complete in terms of host stellar
  mass. The redshift dependent structures that can be seen in this
  figure are due the relatively small size of COSMOS. Our work accounts
  for sample variance using a a suite of mock catalogs.}
\label{agn_sample}
\end{center}
\end{figure}

The conclusions drawn in this paper are specific to the AGN sample
described above. In particular, we do not probe all AGN down to our
mass limit of $\log_{10}(M_*)>10.5$. Because of the lower limit
imposed on $L_X$, our sample will miss AGN with low Eddington ratios
\citep[][]{Aird:2012}. In total, our sample contains 382 AGN with a
mean redshift of $\langle z \rangle=0.7 $, a mean-log X-ray luminosity
of $\langle \log_{10}(L_X) \rangle = 42.7$, and a mean stellar mass of
$\langle M_* \rangle=1.3\times10^{11}$ M$_\odot$. Figure
\ref{agn_sample} shows the $L_X$ and $M_*$ distributions for our
sample.

\subsection{Stellar masses}\label{masses}

In this paper, we use the stellar mass dependent SHMR models and mock
catalogs from L12. For consistency, we adopt the same galaxy stellar
mass estimates as derived in L12. Contamination of the optical light
by emission from the AGN is a potential issue since our masses were
derived using galaxy templates without an AGN component. However, this
effect should be small -- our sample of AGN have moderate luminosities
($L_{\rm x, full} <10^{43.5}$ erg s$^{-1}$) and are not powerful
enough to significantly affect the optical light of the host galaxy
\citep[][]{Nandra:2007,Bundy:2008}. To test for contamination effects,
we compare our mass estimates with those from
\citet[][]{Bongiorno:2012} which were derived using both galaxy and
AGN templates. We find an overall offset of $0.18$ dex between our
mass estimates which is within the expected range of systematic
uncertainties \citep[][]{Behroozi:2010}. More importantly, however,
this mass offset does not exhibit any trends with $L_X$ suggesting
that our mass estimates are robust at these moderate luminosities.

Here we only give a brief description of the stellar mass estimates
and refer the reader to L12 and \citet{Bundy:2010} for further
details. Stellar mass estimates are based on PSF-matched 3\farcs0
diameter aperture photometry from the ground-based COSMOS catalogs
(filters $u^*, B_J, V_J, g^+, r^+, i^+, z^+, K_s$)
\citep[][]{Capak:2007,Ilbert:2009, McCracken:2010}. The depth in all
bands reaches at least 25th magnitude (AB) with the $K_s$-band limited
to $K_s < 24$. Stellar masses are derived using the Bayesian code
described in \citet{Bundy:2006} assuming a Chabrier IMF and a
\citet[][]{Charlot:2000} dust model. An observed galaxy's spectral
energy distribution (SED) and redshift is referenced to a grid of
models constructed using the \citet{bruzual:2003} synthesis code.  The
grid includes models that vary in age, star formation history, dust
content, and metallicity. At each grid point, the probability that the
observed SED fits the model is calculated, and the corresponding
stellar mass to K-band luminosity ratio and stellar mass is stored.
By marginalizing over all parameters in the grid, the stellar mass
probability distribution is obtained.  The median of this distribution
is taken as the stellar mass estimate.

\subsection{Weak Lensing Catalog}\label{wlcatalog}

The COSMOS program has imaged the largest contiguous area (1.64
degrees$^2$) to date with the {\it Hubble Space Telescope (HST)} using
the {\it Advanced Camera for Surveys (ACS)} {\it Wide Field Channel
  (WFC)} \citep[][]{Koekemoer:2007}. The imaging quality of ACS and
the stability of the HST PSF makes this a prime data-set with which to
perform weak lensing measurements at $z<1$. The details of the COSMOS
weak lensing catalog are already described in detail elsewhere
\citep[][]{Leauthaud:2007,Rhodes:2007,Massey:2007,Leauthaud:2012a}. The
COSMOS weak lensing catalog contains $3.9 \times 10^5$ galaxies with
accurate shape measurements which represents a number density of $66$
source galaxies per arc-minute$^{2}$.

The galaxy-galaxy lensing signals presented in section
\ref{wlmeasurements} are measured following the same methodology as
L12. The only minor difference compared to L12 is that here we use an
updated version of the COSMOS photoz catalog (v1.8) of the photometric
redshifts (hereafter photo-$z$'s) presented in \citet{Ilbert:2009}
which have been computed with over 30 bands of multi-wavelength
data. This update to the photo-z catalog does not affect any of the
lensing results.

\subsection{Mock Catalogs}\label{mocks}

The COSMOS ACS survey covers a relatively small volume. To estimate
sample variance, we use a series of mock catalogs described in L11 and
L12. These mocks are extracted from a 1400$^3$ particle, 420 $h^{-1}$
Mpc N-body simulation (``Consuelo'' from the Las Damas suite) with a
particle mass of 1.87$\times 10^{9}$ $h^{-1}$ M$_{\odot}$\footnote{In
  this paragraph, numbers are quoted for $H_0=100$ h
  km~s$^{-1}$~Mpc$^{-1}$}\footnote{{ \small
    http://lss.phy.vanderbilt.edu/lasdamas/simulations.html}} (McBride
et al. in prep). In this paper, we use 100 mock catalogs that are
created from from random lines of sight through the simulation volume
for three redshift intervals: $z_1=[0.22,0.48]$, $z_2=[0.48,0.74]$,
and $z_3=[0.74,1]$. Mocks are populated with galaxies using the
stellar-to-halo mass (SHMR) HOD model of L12. By design, this suite of
mock catalogs matches the the stellar-mass dependent clustering and
galaxy-galaxy weak lensing of COSMOS galaxies from $0.2<z<1.0$. The
mocks are largely complete in terms of stellar mass for the
$\log_{10}(M_*)>10.5$ sample considered in this paper.  Mock galaxies
have stellar masses, redshifts, halo masses, and a central/satellite
identification flag.

\section{Methodology}\label{method}

We begin with an outline of the rationale underlying our
investigation. Our goal is to clearly sketch out the steps in our
proposed methodology so that they may be easily followed by future
studies. Although we focus here on a sample of moderate luminosity
AGN, our methodology can be applied to any sub-population with
stellar mass measurements.

Our approach begins with the assumption that AGN can be described by
the same SHMR as the overall galaxy population. Here we use a SHMR
parametrized as a function of stellar mass, but one could consider additional parameters, such as galaxy color
\citep[\eg,][]{Tinker:2013}. The details of the SHMR-based
model\footnote{The L12 model uses a SHMR for central galaxies and an
  HOD-based prescription for satellite galaxies. For convenience,
  throughout this paper, we refer to the combined model (for centrals
  and satellites) as our ``fiducial SHMR model'', even though
  technically speaking, the SHMR only refers to central galaxies.}
that we use and how it is constrained from COSMOS data are described
in \citet{Leauthaud:2011} and \citet{Leauthaud:2012}. Other models
based on the conditional stellar mass function or abundance matching
techniques would also our purpose
\citep[\eg][]{Yang:2008,Moster:2010,Behroozi:2010,Hearin:2013}.

The first step in our methodology is to choose a statistic (or
multiple statistics) to test the assumption that AGN can be described
by the same SHMR as the overall galaxy population. In this paper we
use galaxy-galaxy lensing but our method can be applied to other
statistics such as the AGN auto-correlation function, or
cross-correlations between AGN and galaxies (ideally binned by stellar
mass).

After computing the statistic of interest, the second step is to
compare the results of this measurement with the prediction from the
fiducial SHMR-based model. The goal of this step is to perform a
null-test of whether AGN populate dark matter halos in the same
fashion as the overall galaxy sample. Predictions from this model can
be computed both analytically or from mock catalogs. Here we mainly
rely on mock catalogs to generate our predictions -- these have the
added advantage of providing an estimate of the sample variance.

An important point to stress here is that when performing this
null-test, ideally the AGN sample should be independent from the
sample used to derive the fiducial SHMR. However, the L12 SHMR model
was derived using all galaxies in the COSMOS field, including the
sub-set of AGN hosts considered here. A better approach would be to
use one half of the COSMOS survey to derive the SHMR and the second
half to compute AGN host statistics. Certainly, this type of approach
can be easily adopted in future large area surveys which will have
more than ample statistical constraining power. In our case, however,
the AGN sample only represents $\sim3\%$ of the galaxy population with
$\log_{10}(M_*)>10.5$ and should only have a minor impact on the
overall SHMR.

A negative null-test would be highly interesting and would indicate
that AGN (or more generally, the sub-population in question) ``know''
something about the dark matter halos in which they reside. In this
case, step three is to vary a sub-set of parameters (those we expect
might be different for active populations). This choice can be
informed by predictions from semi-analytic models of galaxy formation
(SAM) or from direct hydrodynamic simulations
\citep[\eg][]{DeGraf:2012,Chatterjee:2012}. As discussed in more
detail in section \ref{sam1}, one parameter to consider is the AGN
satellite fraction $f_{\rm sat}$. Another parameter of interest might
be $c_{\rm sat}$, the halo concentration of satellite AGN
\citep[e.g,][]{Chatterjee:2012}. Step three is to vary a small set of
parameters (e.g., $f_{\rm sat}$ and/or $c_{\rm sat}$) to fit the
statistic of choice while marginalizing over other parameters in the
SHMR-based model. In this paper, however, step three is unnecessary
because the null-test is positive (see section \ref{wlsignal}).

The final step in our methodology is to use the fiducial SHMR (or the
modified version from step three) to study halo distributions,
satellite fractions, and halo occupation statistics. Again, this step
can be achieved both analytically or by using mock catalogs. This
final step combines two key sets of information. These are: a) the
fiducial (or modified) SHMR-based model and b) the AGN fraction as a
function of stellar mass and redshift.


\section{Results}\label{results}

\subsection{Galaxy-Galaxy Lensing of X-ray AGN}\label{wlmeasurements}

To obtain high signal-to-noise measurements, we stack the weak lensing
signal around our sample of 382 AGN hosts as a function of radial
transverse separation $r$. All of our stacks are performed in physical
coordinates. The galaxy-galaxy lensing signal that we measure yields
an estimate of the mean {\em surface mass density contrast} profile
for our AGN host sample:

\begin{equation}
  \Delta \Sigma(r)\equiv\overline{\Sigma}(< r)-\overline{\Sigma}(r)
\label{dsigma}
\end{equation}

Here, $\overline{\Sigma}(r)$ is the azimuthally averaged and projected
surface mass density at radius r and $\overline{\Sigma}(< r)$ is the
mean projected surface mass density within radius r
\citep[e.g,][]{Miralda-Escude:1991,Wilson:2001}. For the radial ranges
that we probe in this study ($<$ 2 Mpc), our lensing signals are
mainly due to the dark matter halos associated with the stacked galaxy
sample (the ``one-halo'' term).

Uncertainties on the lensing signal are derived using two different
methods. The first, most naive estimate assumes that the data bins are
independent, and that measurement error and shape noise are the
dominant sources of error. The uncertainty on $\Delta\Sigma$ is then
simply $\sigma_{w}=\sqrt{1/\sum w_i}$ where the sum is performed over
all lens-source pairs and where $w_i$ is an estimate of the shear
variance for each source (see L12). However, at larger radii, bins may
become correlated due to the fact that the same source galaxy may be
associated with multiple lens galaxies (``correlated shape
noise''). To test for the magnitude of this effect, we also derive
jack-knife uncertainties on $\Delta\Sigma$, noted hereafter as
$\sigma_{jk}$. The two uncertainty estimates are in good agreement
with the jack-knife errors being somewhat larger for the outer radial
bins suggesting small levels of correlated shape noise. Jack-knife
estimates of covariance for the outer radial bins suggest that the
correlation coefficient between the three last radial bins is at most
$|\rho|<0.3$. Throughout this paper we quote values using both of
these uncertainty estimates and we neglect the small amount of
covariance for the outermost radial bins. Finally, we use 100 mock
mock catalogs (described in the previous section) to estimate the
sampling variance for our lensing signal. These include both shot
noise due to the small number of lens galaxies in our sample, as well
as sample variance in the underlying dark matter realization for a
field the size of COSMOS. These errors are sub-dominant (less than
10\%) compared to shape noise.

\begin{figure*}
\begin{center}
\includegraphics[width=17cm]{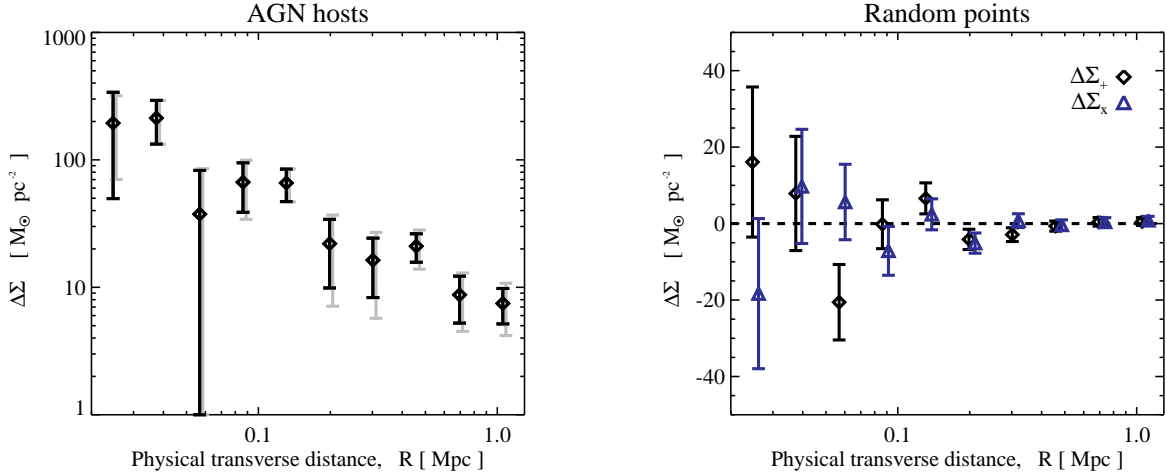}
\caption{{\it Left}: First reported weak lensing signal of X-ray AGN
  as measured from 382 X-ray selected hosts with $\langle \log(L_X)
  \rangle = 42.7$ from the COSMOS field. Black error bars show
  shape-noise uncertainties, grey error bars indicate jack-knife
  uncertainties. {\it Right}: As a test for systematics, we also
  compute the lensing signal measured around random points.}
\label{agn_lensing}
\end{center}
\end{figure*}

Our weak lensing signal for the AGN sample is shown in Figure
\ref{agn_lensing}. For this measurement, we have used 10
logarithmically spaced bins from $r=20$ kpc to $r=1.3$ Mpc. The weak
lensing signal is clearly detected out to the largest scales with a
mean signal-to-noise per data point of $S/N\sim 2.4$ using shape noise
uncertainties and $S/N\sim 2.1$ using jack-knife uncertainties. As a
test for systematic effects, we also compute the lensing signal around
7000 random points that are drawn from the same redshift distribution
as our AGN lens sample. The result is shown in the right hand panel of
Figure \ref{agn_lensing}. No evidence for systematic shear patterns
are detected around random points.

\subsection{Weak Lensing Signal of AGN Hosts Compared to Fiducial
  Stellar-to-Halo Mass Model}\label{wlsignal}

\begin{figure*}
\begin{center}
\includegraphics[width=16cm]{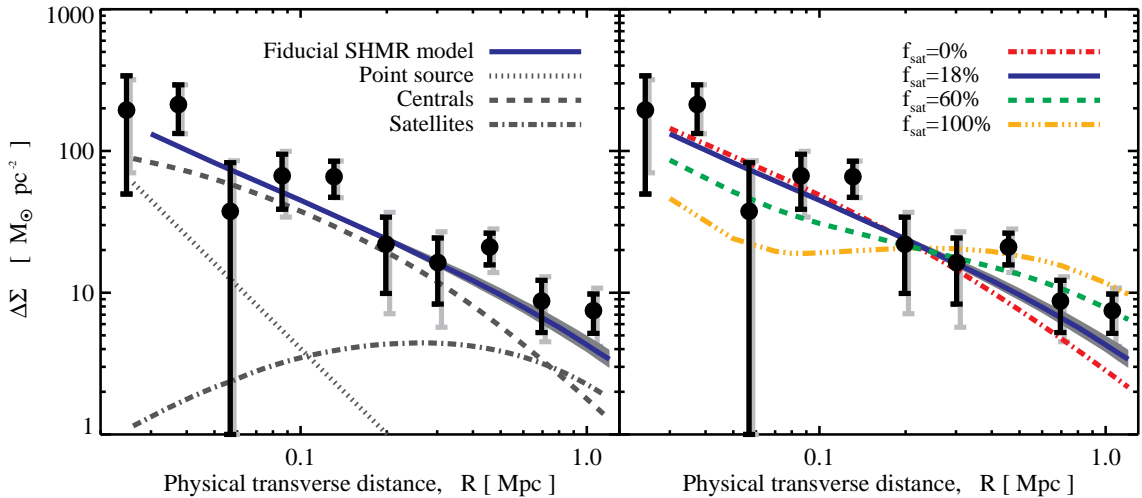}
\caption{{\it Left}: Lensing signal of AGN sample compared to the
  prediction from our fiducial SHMR model (blue line). The grey shaded
  region shows the field-to-field variance derived from 100 mock
  catalogs. These uncertainties are insignificant compared to shape
  noise uncertainties. The lensing signal has contributions from the
  host stellar mass (dotted line) and from the dark matter halos
  associated with both central (dashed line) and satellite (dash-dot
  line) galaxies. We confirm the null hypothesis that the AGN host
  occupation is no different than that defined by galaxies with the
  same $M_*$, regardless of nuclear activity. {\it Right}: predicted
  lensing signal for varying satellite fractions. The satellite
  fraction of the fiducial SHMR model is $f_{\rm sat}=18\%$.}
\label{agnmocks}
\end{center}
\end{figure*}

Given the host mass and redshift of each AGN in our sample, we use the
SHMR model of L12 to compute the predicted AGN galaxy-galaxy lensing
signal. The predicted galaxy-galaxy lensing signal is shown in Figure
\ref{agnmocks} and is composed of three terms: 1) a contribution from
the stellar mass of the AGN host galaxy, 2) a contribution from the
dark matter halos associated with central galaxies that follows the
the standard Navarro-Frenk-White profile \citep[NFW;][]{Navarro:1997},
3) and a contribution from the dark matter halos associated with
satellite galaxies. The total weak lensing signal is the sum of these
three terms. Contributions from sub-halos associated with satellites
are neglected. The ``two-halo'' term is negligible on these small
radial scales.

The grey shaded region in Figure \ref{agnmocks} shows the
field-to-field variance derived from mock catalogs; this is
sub-dominant compared to the measurement errors on the lensing
signal. Overall, we find that our fiducial SHMR model does an
excellent job at matching the weak lensing signal of AGN. The $\chi^2$
between the measured lensing signal and the SHMR prediction is
$\chi^2/d.o.f=8.5/10$ ($\chi^2=11.6/10$ for shape-noise errors). Since
there are no free parameters in this model, the number of degrees of
freedom is simply the number of data points, $d.o.f=10$. As mentioned
in section \ref{method}, however, the AGN sample is not strictly
independent from the data used to infer the SHMR which does place a
caveat on this comparison.

Our null-test is positive. Hence, we confirm the null hypothesis that
the AGN host occupation is no different than that defined by galaxies
with the same $M_*$, regardless of nuclear activity.

Step three in our methodology in unnecessary for this sample
(we do not need to vary any parameters to describe the lensing
signal). As an example, however, of how step three might proceed - the
right hand panel of Figure \ref{agnmocks} shows how the predicted
lensing signal of AGN hosts varies with $f_{\rm sat}$ (keeping all
other parameters fixed). We find that reducing the satellite fraction
to $f_{\rm sat}=0$ only has a relatively small impact on the overall
lensing signal. The predicted lensing signal is mostly un-changed on
small scales and decreases slightly on 1 Mpc scales. If we increase
the satellite fraction to 100\% then the predicted signal increases on
large scales but decreases on small scales creating a clear
scale-dependent signature which should be easily detectable with the
next generation of lensing surveys. Small values of $f_{\rm sat}$ may
be difficult to detect with lensing alone, but joint measurements of
lensing and clustering will be able to pin down $f_{\rm sat}$ with
greater accuracy.

\subsection{Dark Matter Environment of AGN sample as Inferred from
  Host Mass}\label{dmprediction}

In the previous section, we compared the weak lensing signal of AGN
hosts with the prediction from our fiducial SHMR model. The fact that
they are statistically indistinguishable suggests that AGN in our
sample populate halos in the same fashion as the overall galaxy
population. One caveat, however, is that our AGN lensing signal is
relatively noisy. Upcoming lensing surveys with better signal-to-noise
may find differences that we have been unable to detect. In the
meantime, given that we have no evidence to suggest otherwise, in the
remainder of this paper we proceed under the assumption that host
stellar mass and redshift are sufficient to predict the mean dark
matter environment for this AGN sample.

\begin{figure}
\begin{center}
\includegraphics[width=8.9cm]{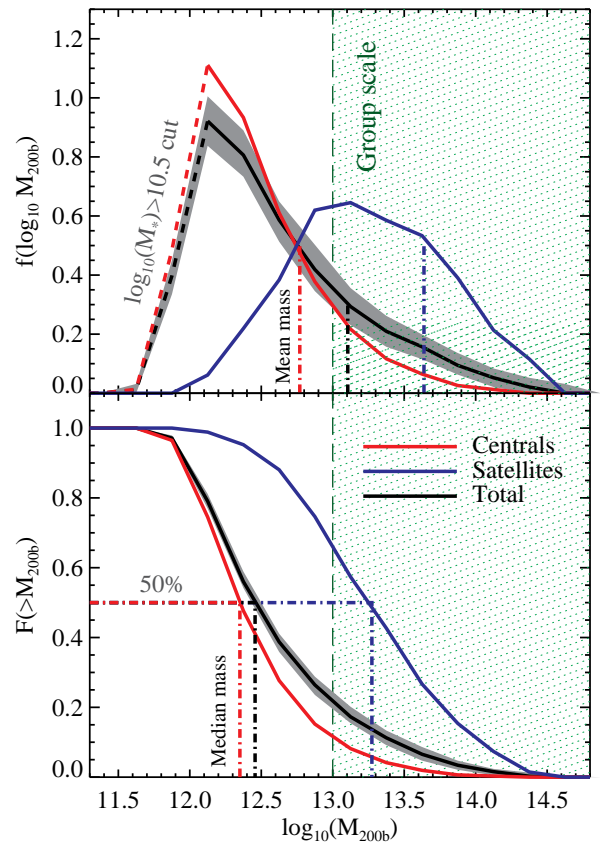}
\caption{{\it Top}: probability per $\log_{10}(M_{\rm 200b})$ that an AGN
  in our sample is hosted by a halo of mass $M_{\rm 200b}$. Note that this
  is a probability density function and may take on values greater
  than 1. Distributions are shown separately for central (red line)
  and satellite AGN (blue line). For satellites, $M_{\rm 200b}$ represents
  the mass of the parent halo (not the sub-halo). The black solid line
  is the full sample (centrals plus satellites) with grey shaded
  regions representing the variance from mock catalogs. Vertical
  lines (dash-dot) indicate a mean halo mass. The dashed green
  vertical line shows a typical mass limit for galaxy groups. Note
  that the sharp drop-off at $\log_{10}(M_{\rm 200b})\sim 12$ is simply
  due to the fact that we select AGN with hosts mass
  $\log_{10}(M_*)>10.5$. This cut was motivated to select a sample
  that is roughly similar to previous work on the clustering of X-ray
  AGN. The black and red curves would continue to rise had we included
  lower mass AGN hosts in our sample. {\it Bottom}: Complementary
  cumulative distribution function. Vertical lines (dash-dot) indicate
  a median halo mass. Only $\sim$60\% of AGN satellites are contained in
  halos with $\log_{10}(M_{\rm 200b})>13$.}
\label{agn_halo_masses}
\end{center}
\end{figure}

We now use our mock catalogs to investigate the predicted halo mass
distribution for this AGN sample. A mock AGN population is extracted
from each mock catalog (see section \ref{mocks}) by matching mock
galaxies and AGN hosts in terms of stellar mass and redshift. There
are 100 mock catalogs in total, each mock has the same volume as
COSMOS. Figure \ref{agn_halo_masses} shows the halo mass probability
density function as well as the complementary cumulative distribution
function for mock AGN samples. Errors in Figure \ref{agn_halo_masses}
represent the field-to-field variance between mock catalogs. Table
\ref{mh_table} summarizes the mean and median halo masses for
centrals, satellites, and for the combined sample (centrals and
satellites). For the combined sample, we find that the mean halo mass,
$\langle M_{\rm 200b} \rangle=1.3\times$10$^{13}$ M$_{\odot}$, is a
factor of 4.5 times larger than the median halo mass, $M_{\rm
  200b}^{\rm med}=2.9\times$10$^{12}$ M$_{\odot}$. We underscore the
fact that the mean and the median halo masses may be markedly
different because of the skewed tail in the halo mass distribution.

\begin{table}
\caption{Mean and median halo masses} \label{mh_table}
\begin{tabular}{@{}lcc}
\hline
Halo mass & Mean $\langle M_{\rm 200b} \rangle$&  Median $ M_{\rm 200b}^{\rm med}$\\ 
   & (10$^{13}$ M$_{\odot}$)&  (10$^{13}$ M$_{\odot}$)\\ 
\hline
Centrals & 0.59$\pm$0.08&  0.22$\pm$0.01 \\ 
Satellites & 4.3$\pm$1.3 &  1.9$\pm$0.5 \\ 
Cen + Sat & 1.3$\pm$0.3  &  0.29$\pm$0.02 \\ 
  \hline
 \end{tabular}

\medskip
Note: errors represent the field-to-field variance derived from mock catalogs.
\end{table}

We stress that these values are specific to our particular AGN sample
selection. In our case, the most important factor in determining the
exact halo mass distribution is the $\log_{10}(M_*)>10.5$ cut that we
applied to the AGN host masses. This cut drives the sharp drop-off at
$\log_{10}(M_{\rm 200b})\sim 12$ in Figure \ref{agn_halo_masses}. However,
in practice, COSMOS AGN catalogs do contain X-ray AGN in galaxies with
$\log_{10}(M_*)<10.5$. According to the SHMR, on average, these are
expected to live in even lower mass halos.

\begin{figure}
\begin{center}
\includegraphics[width=8.5cm]{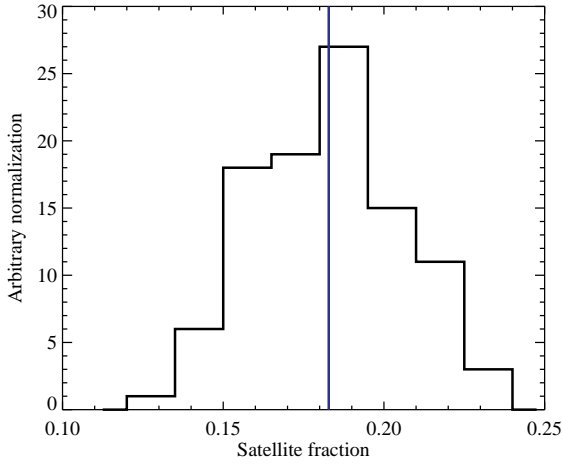}
\caption{AGN satellite fraction distribution from mock catalogs. The
  mean satellite fraction is $\langle f_{\rm sat} \rangle =18\%$ with a rms dispersion
  between mock catalogs of 2\%.}
\label{agn_sat}
\end{center}
\end{figure}

Let us now turn our attention to the halo mass distributions of
satellite AGN\footnote{For satellites, halo mass refers to the mass of
  the parent halo, not sub-halo masses.}. Our predictions are based on
our fiducial SHMR model where AGN hosts have the same satellite
fractions as inactive galaxies (see section \ref{wlsignal} and Figure
\ref{agnmocks}). We find that 50\% of satellite AGN in our sample live
in halos less massive than $\log_{10}(M_{\rm 200b})=13.2$. Figure
\ref{agn_sat} shows the predicted satellite fractions for our
sample. We find a mean satellite fraction of $\langle f_{\rm
  sat}\rangle =18\%$ with a rms dispersion between mock catalogs of
2\%. How does this compare with previous results derived from
clustering studies of X-ray AGN? Reliable constraints on satellite
fractions derived from HOD modeling are limited by modeling
uncertainties \citep[\eg][]{Miyaji:2011,Shen:2013}. A perhaps more
robust estimate of satellite fractions may be obtained by measuring
the effects of satellite peculiar velocities on the 2d redshift space
correlation function. Using this technique, \citet{Starikova:2011}
report a 90\% confidence level upper limit on the satellite fraction
of $f_{\rm sat}<8\%$. Their sample, however, is truncated to brighter
hosts than ours for which we do indeed expect lower satellite
fractions. As discussed in more detail in section \ref{discussion}, it
is unclear how much this difference might be of genuine interest as
opposed to simply due to sample selection effects.

\subsection{Halo Occupation Functions}\label{hod}

\begin{figure*}
\begin{center}
\includegraphics[width=16cm]{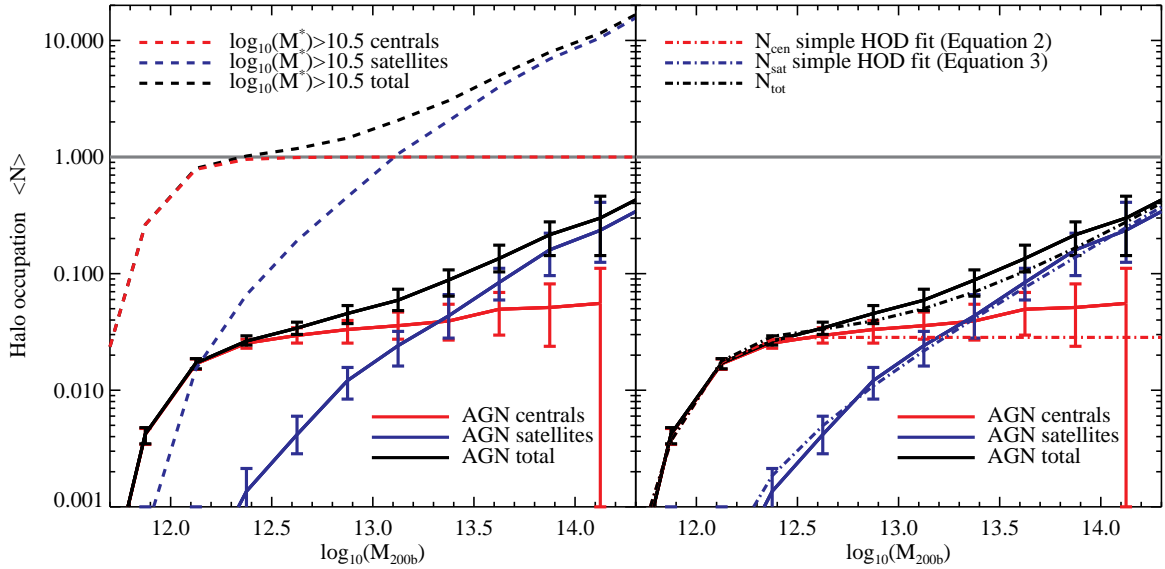}
\caption{{\it Left}: Central and satellite halo occupation functions
  for our AGN sample derived from mock catalogs based on our fiducial
  SHMR model (solid lines). Errors represent field-to-field variance
  for a COSMOS like survey. Dashed lines indicate the occupation
  functions for all galaxies with $\log_{10}(M_*)>10.5$. The turn-over
  in the HOD at $\log(M_{\rm 2000b})\sim12$ is set by the
  $\log_{10}(M_*)>10.5$ cut that defines our AGN sample. The amplitude
  of the AGN HOD ($\langle N_{\rm cen}\rangle \sim$0.01 - 0.06)
  indicates that X-ray obscured AGN from our sample only represent a
  few percent of all central galaxies in any given halo mass bin. {\it
    Right}. Comparison between the occupation function inferred from
  our analysis and a fit using a simple HOD given by Equations
  \ref{ncen} and \ref{nsat}.}
\label{agn_hod}
\end{center}
\end{figure*}

The occupation functions for AGN in our sample are shown in Figure
\ref{agn_hod}. For reference, we also show the occupation functions
for a galaxy sample with $\log_{10}(M^*)>10.5$. We find that
$\log_{10}(M_{\rm 200b}) \sim 12$ halos will host on average $\langle
N_{\rm cen}\rangle \sim$0.01 central AGN in our sample and at
$\log_{10}(M_{\rm 200b}) \sim 14$ this number rises to $\langle N_{\rm
  cen}\rangle \sim$0.06 (see Table \ref{hod_table}). Including both
centrals and satellites, we expect that $\log_{10}(M_{\rm 200b}) \sim
14.0$ halos host on average $\langle N_{\rm tot}\rangle \sim$0.3 AGN
in our sample. The errors on the occupation functions due to
field-to-field variance are non negligible for groups with mass
$\log_{10}(M_{\rm 200b}) \sim 14.0$ for a survey the size of COSMOS.

\begin{table}
\caption{Halo occupation functions} \label{hod_table}
\begin{tabular}{@{}cccc}
\hline
$\log_{10}(M_{\rm 200b})$ & $\langle N_{\rm cen} \rangle$ & $\langle
N_{\rm sat} \rangle$ & $\langle N_{\rm tot} \rangle$ \\
\hline
 11.875& $0.0043_{-0.0007}^{+0.0005}$& $0.0$ &$0.0042_{-0.0008}^{+0.0006 }$\\
 12.125& $0.017_{-0.002}^{+0.002}$&$0.0$  &$0.017_{-0.002 }^{+0.002 }$\\
 12.375& $0.025_{-0.002}^{+0.002}$&$0.0014_{-0.0007}^{+0.0008 }$ &$0.026 _{-0.002 }^{+0.003 }$\\
 12.625& $0.028_{-0.004}^{+0.005}$& $0.0041_{-0.001}^{+0.002 }$  &$0.034_{-0.004 }^{+0.004}$\\
 12.875& $0.033_{-0.008}^{+0.006}$&$0.012_{-0.004 }^{+0.004 }$  &$0.046_{-0.008 }^{+0.008 }$\\
 13.125& $0.036_{-0.008}^{+0.01}$&$0.024 _{-0.008}^{+0.008 }$  &$0.059_{-0.01}^{+0.01 }$\\
 13.375& $0.039_{-0.01}^{+0.02}$&$0.04_{-0.02 }^{+0.02}$   &$0.088_{-0.02}^{+0.02 }$\\
 13.625& $0.05_{-0.02}^{+0.02}$&$0.08_{-0.02}^{+0.03 }$    &$0.13_{-0.03}^{+0.04 }$\\
 13.875& $0.05_{-0.03}^{+0.03}$&$0.16_{-0.06 }^{+0.06 }$   &$0.22_{-0.07}^{+0.06 }$\\
 14.125& $0.06_{-0.06}^{+0.05}$& $0.2_{-0.1 }^{+0.2 }$    &$0.3_{-0.2}^{+0.2 }$\\
  \hline
 \end{tabular}
\end{table}

Our occupation functions are mostly comparable to those obtained by
\citet{Allevato:2012} from direct counting of AGN in groups within the
COSMOS field. \citet{Allevato:2012} measure $\langle N_{\rm
  tot}\rangle \sim$0.2-0.6 \footnote{\citet{Allevato:2012} correct
  their HODs for incompleteness in $L_X$ but we attempt no such
  corrections here. The values quoted here from \citet{Allevato:2012}
  are taken from their Figure 3 before any luminosity and redshift
  evolution corrections} for halos with masses above $10^{13}$
M$_{\odot}$. Our values are in fair agreement with these estimates,
especially given that we apply different selection criteria to the
COSMOS AGN samples (we apply host mass and X-ray luminosity cuts for
example) which can easily translate into factors of 2 differences in
the amplitude of the inferred HODs.

We stress that the goal of this paper is not so much the exact values
of the HOD presented in Figure \ref{agn_hod} since these will depend
sensitively on our particular AGN sample selection (varying the $L_X$
cuts will affect the amplitude of the HOD for example). Instead, our
main point here is to demonstrate that, under the assumption that
active and inactive galaxies inhabit similar dark matter
environments, the SHMR-based approach advocated here makes firm
predictions for the {\em shape and normalization} of the AGN
occupation functions. \citet{Miyaji:2011} investigated three different
HOD parameterizations to model the cross-correlation function between
ROSAT All-Sky Survey detected AGN and SDSS LRGs. Among the three
models explored by \citet{Miyaji:2011}, their model B provides the
best description of the HODs presented here. This is a model that is
similar to those used for threshold galaxy samples but with an
additional free parameter $f_A$ that allows the global normalization
of $\langle N_{\rm cen} \rangle$ to float.

Figure \ref{agn_hod} shows that our HOD is reasonably well fit by an HOD
of the form

\begin{equation}\label{ncen}
\langle N_{\rm cen} \rangle = \frac{f_A}{2}\left[
  1+\mbox{erf}\left(\frac{\log_{10}( M_{\rm 200b}/M_{\rm min})}{\sigma_{logM}}\right)\right]
\end{equation}

\begin{equation}\label{nsat}
\langle N_{\rm sat} \rangle = \langle N_{\rm cen} \rangle \left(\frac{M_{\rm 200b}}{M_{\rm sat}}\right)^{\alpha} \exp\left(\frac{-M_{\rm cut}}{M_{\rm 200b}}\right)
\end{equation}

\noindent with $f_A=0.028$, $M_{\rm min}=1.2\times$10$^{12}$
M$_{\odot}$, $\sigma_{logM}=0.25$, $M_{\rm sat}=1.5\times$10$^{13}$
M$_{\odot}$, and $M_{\rm cut}=2\times$10$^{12}$ M$_{\odot}$. This
model is similar to model B from \citet{Miyaji:2011} except that
$\langle N_{\rm cen} \rangle$ is modeled with an error function
instead of a step function, and our satellite occupation includes an
exponential cutoff with a scale set by $M_{\rm cut}$. 

One interesting feature in Figure \ref{agn_hod} is that our
empirically determined HODs displays a rise towards higher halo mass
that is not well captured by a constant $f_A$.  This parameter is
sometimes interpreted as an AGN duty cycle
\citep[\eg][]{Martini:2001,Shen:2007, Miyaji:2011,Richardson:2013}. In
this case, the rising nature of $\langle N_{\rm cen} \rangle$ could
indicate a varying AGN duty cycle with halo mass. We caution however
that at least part of this trend will be imposed by sample selections
effects introduced by our $L_X$ cut. AGN show a wide distribution of
Eddington ratios
\citep[e.g.][]{Heckman:2004,Kauffmann:2009,Schulze:2010,Aird:2012}. For
a fixed $L_X$ cut we sample AGN in massive galaxies over a wide range
of Eddington ratios, while for less massive galaxies we only sample
AGN with larger Eddington ratios \citep{Schulze:2010,Aird:2012}. This
may lead to an apparent increase in $f_A$ with halo mass for $L_X$
selected samples.

\section{Discussion}\label{discussion}

\subsection{Comparison with Previous Results Based on Clustering
  Measurements}\label{clus}

Before comparing with previous results, let us first briefly review
how clustering studies typically infer halo mass. What most studies
based on clustering measurements actually derive\footnote{Here we
  refer specifically to studies that infer halo mass from
  $b_{\rm{eff}}$ assuming that
  $b(M_{\rm{eff}})=b_{\rm{eff}}$. Studies that model clustering with
  an HOD type approach may quote a mean, median, or a minimum halo
  mass instead of an effective halo mass.} is the linear effective
bias, $b_{\rm{eff}}$. The effective halo mass is then the mass which
satisfies $b(M_{\rm{eff}})=b_{\rm{eff}}$ where $b(M_{\rm h})$ is the
mean bias of halos of mass $M_{\rm h}$
\citep[\eg][]{Tinker:2010}. What exactly does this effective halo mass
correspond to when considering samples that span a wide range in halo
mass? The effective bias measured by clustering studies is:

\begin{equation}\label{bias_eq}
  b_{\rm{eff}}=\frac{\int b(M_{\rm h})N_{\rm AGN}(M_{\rm h})n(M_{\rm h})\rm{d}M_{\rm h}}{\int N_{\rm AGN}(M_{\rm h})n(M_{\rm h})\rm{d}M_{\rm h}}
\end{equation}

\noindent where $N_{\rm AGN}(M_{\rm h})$ and $n(M_{\rm h})$ are respectively the mean
number of AGN and the number density of halos as a function of $M_{\rm h}$
\citep[\eg][]{Baugh:1999,Fanidakis:2013}. For our purpose, it is
perhaps more clear to re-write Equation \ref{bias_eq} so as to
highlight the AGN halo mass probability density function $f_{\rm
  AGN}$:

\begin{equation}\label{fagn}
  f_{\rm AGN}=\frac{ N_{\rm AGN}(M_{\rm h})n(M_{\rm h})}{\int N_{\rm AGN}(M_{\rm h})n(M_{\rm h})\rm{d}M_{\rm h}}
\end{equation}

Using $f_{\rm AGN}$, Equation \ref{bias_eq} simply becomes:

\begin{equation}\label{bias_eq2}
  b_{\rm{eff}}=\int b(M_{\rm h}) f_{\rm AGN}(M_{\rm h})  \rm{d}M_{\rm h}
\end{equation}

Written in this fashion, it is clear that $M_{\rm{eff}}$ measured
from $b_{\rm{eff}}$ corresponds to a bias-weighted average of $f_{\rm
  AGN}$. Halo bias $b(M_{\rm h})$ is not a simple linear function of halo
mass. Broadly speaking, $b(M_{\rm h})$ is a shallow function at low halo
mass and then rises sharply at higher halo mass
\citep[\eg][]{Tinker:2010}. Hence, $M_{\rm{eff}}$ may be different
than other, perhaps more useful averages such as the mode, median, or
mean value of $f_{\rm AGN}$.

We now investigate the difference between $M_{\rm{eff}}$, the median,
and the mean of our halo mass distribution. At our mean redshift of
$\overline{z}=0.7$, our SHMR model predicts $b_{\rm{eff}}=1.8$ and
$M_{\rm 200b}^{\rm eff}=5.0\times$10$^{12}$ M$_{\odot}$. Hence,
$M_{\rm 200b}^{\rm eff}$ is roughly mid-way between the median and the
mean halo mass of $f_{\rm AGN}$.

\begin{figure*}
\begin{center}
\includegraphics[width=17.5cm]{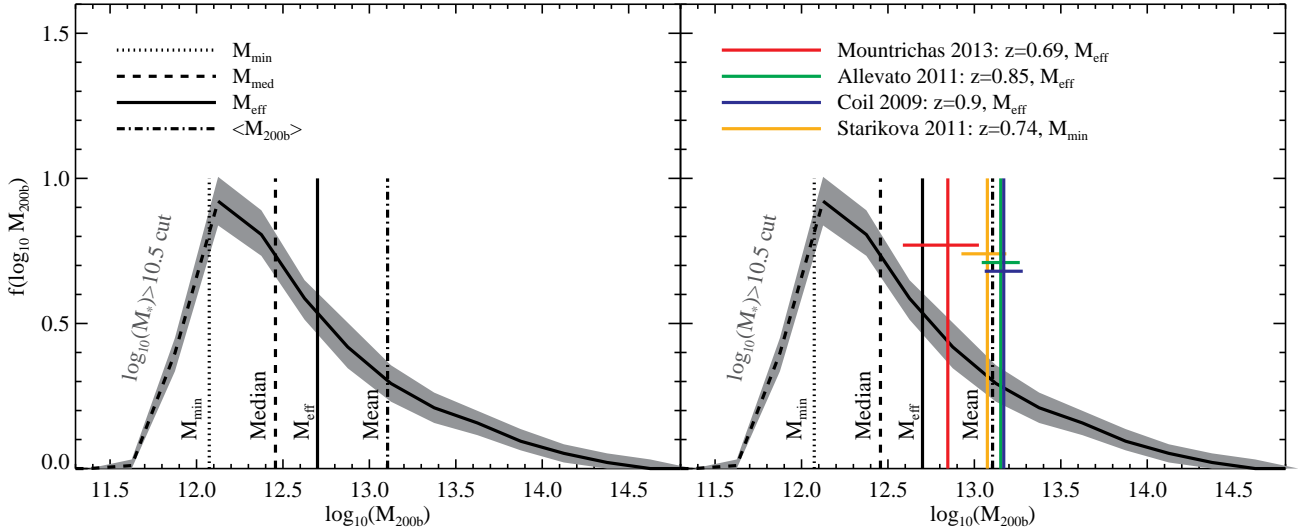}
\caption{{\it Left}: probability per $\log_{10}(M_{\rm 200b})$ that an
  AGN in our sample is hosted by a halo of mass $M_{\rm 200b}$. The
  dotted, dashed, solid and dash-dot vertical lines show $M_{\rm min}$
  (Equation \ref{ncen}), $M_{\rm{med}}$, $M_{\rm{eff}}$ and $\langle
  M_{\rm 200b} \rangle$ respectively. {\it Right}: comparison with
  with values inferred from studies of X-ray AGN clustering. Previous
  results from clustering studies tend to lie in-between our effective
  halo mass and our mean halo mass.}
\label{agn_halo_masses_comapre}
\end{center}
\end{figure*}

Figure \ref{agn_halo_masses_comapre} compares our halo mass
distribution with $M_{\rm{eff}}$ values derived from clustering
studies. We focus on samples that (only very roughly) span the same
redshift and luminosity range as ours \citep[][]{Coil:2009,
  Allevato:2011,Mountrichas:2013}\footnote{Unless stated otherwise, we
  used a compilation of halo mass values provided in Table 2 of
  \citet[][]{Fanidakis:2013}. For \citet{Allevato:2011}, we use their
  results from Table 3 of their paper for obscured X-ray AGN at
  $\overline z=0.85$. All halo masses have been converted to our mass
  definition, $M_{\rm 200b}$.}. We also compare with the results from
\citet[][]{Starikova:2011} at $0.5<z<1.0$ which are inferred from the
ratio of the projected auto-correlation function integrated along and
across the line of sight. The technique employed in
\citet[][]{Starikova:2011} is different compared to the other studies
mentioned above because it uses additional information from the
peculiar velocities of satellites. 

Figure \ref{agn_halo_masses_comapre} shows that the halo mass values
reported by previous studies are typically larger than ours and tend
to lie in-between our effective halo mass and our mean halo mass. Our
results are most different compared to \citet[][]{Starikova:2011} who
report a {\em minimum} halo mass\footnote{\citet[][]{Starikova:2011}
  quote the halo mass that corresponds to the minimum $V_{max}$
  (maximum circular velocity) of halos that host an X-ray AGN in their
  model.}, not an effective halo mass. However, our host sample
selection is also most different compared to
\citet[][]{Starikova:2011} who limit their selection to brighter hosts
than we do. This brings us to an important point, discussed in the
following section, which is that when comparing studies of AGN
clustering -- any cuts on host galaxy mass/luminosity must be taken
into account because host mass/luminosity {\em correlates} with halo
  mass.

\subsection{A Simple Selection Effect that Cannot be Neglected: Brighter
  Hosts live in More Massive Halos}\label{seleffect}

Studies of X-ray AGN clustering are typically limited to brighter
hosts simply due to the fact that such measurements require
spectroscopic redshifts. The samples we compare with in Figure
\ref{agn_halo_masses_comapre} are typically limited to hosts with
I$_{AB}<$21.5-23. More specifically, \citet[][]{Coil:2009} make no
explicit cut on host luminosity but their sample is roughly limited at
$R_{AB}< 22.6$ due to the availability of optical spectroscopy. The
\citet[][]{Allevato:2011} AGN sample is limited to I$_{AB}<$23 (with a
spectroscopic completeness of 53\%). \citet[][]{Koutoulidis:2013} make
no explicit cut on host luminosity, but spectroscopic requirements
drive an implicit cut on host luminosity which varies between the
different surveys in their compilation. \citet[][]{Starikova:2011}
apply a host magnitude cut of $I<21.5$ to their sample. There is not a
one-to-one relation between stellar mass and optical luminosity, but
to give some idea of the characteristic luminosity of our hosts,
galaxies in COSMOS with $\log_{10}(M_*)>10.5$ have a median magnitude
of $I_{AB}\sim 21$ at $z\sim 0.5$. At $z\sim$1 they have a median
magnitude of $I_{AB}\sim 22.6$.

On average, hosts with brighter luminosities live in larger dark
matter halos. In addition, samples defined by a fixed observed host
luminosity threshold probe different stellar mass (hence halo mass)
limits as a function of redshift. We stress that these (sometimes
implicit) cuts on host properties need to be considered when comparing
X-ray selected samples to one another, when comparing AGN samples selected
at different wavelengths (e.g, X-ray AGN versus QSOs), and also when
comparing with theoretical predictions from SAMs or hydrodynamical
simulations. Differences in AGN luminosities between samples are
commonly taken into account, but cuts on host properties must also be
considered.

\subsection{The Dark Matter Environment of Moderate Luminosity X-ray
  AGN Compared to UV luminous QSOs}

The prevailing wisdom from clustering studies of X-ray AGN is that
moderate luminosity X-ray selected AGN populate group-sized dark
matter halos with $M_{\rm h} \sim 10^{13} M_{\odot}$ \citep[\eg][to
cite a few recent
examples]{Koutoulidis:2013,Fanidakis:2013,Hutsi:2014}. In contrast, UV
luminous QSOs in the 2DF and SDSS surveys are found to reside in lower
mass halos with $M_{\rm h} \sim 10^{12} M_{\odot}$
\citep[][]{Croom:2005,da-Angela:2008,Ross:2009,Shanks:2011}. This
environmental dependance has led to the suggestion that moderate
luminosity X-ray AGN and luminous QSOs may have different fueling
mechanisms \citep[\eg][]{Fanidakis:2013}. In this scenario, QSOs are fueled from cold-gas reservoirs
that are funneled to galaxy centers by catastrophic events such as
mergers or disk instabilities whereas moderate luminosity X-ray AGN
may be connected with an additional fueling channel in which gas is
accreted directly from a diffuse state in massive dark matter halos
($M_{\rm h} > 10^{12}$ M$\odot$, the ``radio'' or ``hot-halo'' mode).

Under closer consideration, however, the difference between the dark
matter environment of moderate luminosity X-ray AGN and QSOs may not
be so clear. First, selection cuts on host properties (see the previous section) must be taken into
account. Spectroscopic requirements impart different selections on
host properties for X-ray and QSOs samples
\citep[\eg][]{Hopkins:2009a} -- this will naturally lead to difference
in the underlying dark matter distributions. Second, clustering
studies often report a single halo mass scale which may be difficult
to interpret in the context of samples that span a wide range of halo
masses. 

The results of this paper favor a different picture for the dark
matter halos of galaxies hosting moderate luminosity X-ray AGN. Figure
\ref{agn_halo_masses} suggests that most AGN in our sample do {\em
  not} live in group environments ($M_{\rm h} > 10^{13}$
M$_{\odot}$)---50\% of the AGN in our sample are found in halos less
massive than $M_{\rm 200b}\sim3\times 10^{12}$ M$_{\odot}$ and hence
live in relatively low-mass halos. Recently, \citet{Conroy:2013}
showed that a simple phenomenological model in which QSOs live in a
wide range of halos masses successfully predicts both the QSO
luminosity function and the two-point correlation function from $0.5 <
z < 2$. Taken together, these two results suggest that both QSOs and
moderate luminosity X-ray AGN may occupy halos in a relatively
``normal'' way compared to galaxies without active nuclei. The notion
that they may share similar dark matter environments calls into
question the need for different physical mechanisms to explain the
fueling of moderate luminosity X-ray AGN and QSOs.

\subsection{Insights from Galacticus SAM: Halo Mass Distributions of Active
  Galaxies versus All Galaxies}\label{sam1}

The approach adopted in this paper is valid if AGN
populations can be described by varying a few simple parameters in the
SHMR description for the overall galaxy population. How does this
premis compare with theoretical models of AGN activity and which
parameters in the SHMR are most likely to differ? To investigate these questions, we turn
to the state-of-the art Galacticus SAM \citep{Benson:2012}. We use
Galacticus because its modeling of BH physics is relatively realistic
compared to other SAMs, comparable to the detailed BH evolution
modeling developed by \cite{Fanidakis:2011}. Specifically, for this
work, we use v0.9.1 (revision 1456) of Galacticus and the default set
of parameters supplied with that version. A description of the key
features of this SAM relevant for this paper is given in the
Appendix. The full details of the Galacticus model can be found in
\citet{Benson:2012}.

Our goal is not a one-to-one comparison between Galacticus
and our COSMOS results. Many aspects of the COSMOS data are not
reproduced by this Galacticus model. For example, as discussed in the
following section, the overall SHMR is different. Also, at fixed stellar mass, satellites in Galacticus
populate more massive halos than in COSMOS. With these caveats in
mind, we use Galacticus to investigate qualitative differences
between active and inactive galaxies that may be informative in
interpreting our observational results.

\begin{figure}
\begin{center}
\includegraphics[width=8cm]{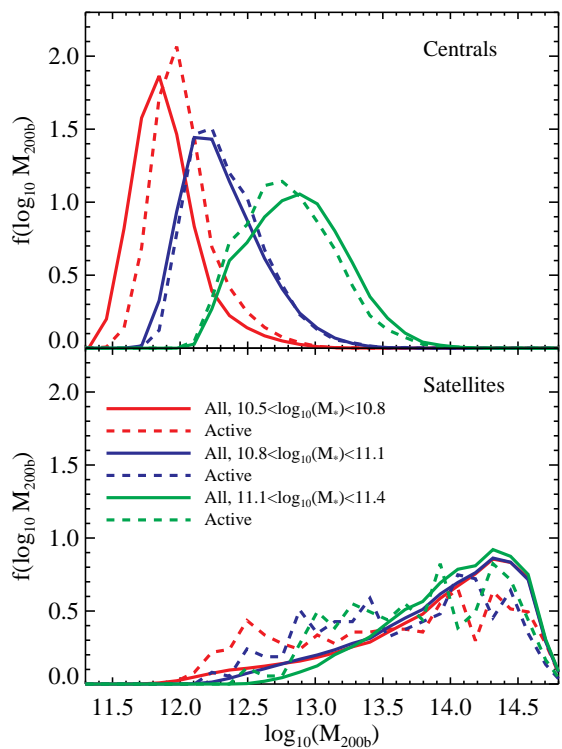}
\caption{Halo mass probability density functions for centrals (upper
  panel) and satellites (lower panel) from the Galacticus SAM. Solid
  lines correspond to all galaxies in three stellar mass bins spanning
  the range $\log_{10}(M_*)=10.5$ to $\log_{10}(M_*)=11.4$. Dashed
  lines correspond to a sample of active galaxies selected to roughly
  mimic our COSMOS sample. Broadly speaking, active galaxies have
  similar halo mass distributions compared to inactive galaxies of
  similar stellar mass.}
\label{galacticus_mhalo_dis}
\end{center}
\end{figure}

We select a sample of active galaxies from the Galacticus simulation
at $z=0.61$ (close to our mean redshift of $z=0.7$) using the same
host mass and $L_X$ cuts as our COSMOS AGN sample. We do not however
mimic the X-ray luminosity incompleteness in the COSMOS data. Figure
\ref{galacticus_mhalo_dis} shows the halo mass distributions of active
galaxies in three stellar mass bins compared to halo mass
distributions for the overall galaxy population. There are some small
differences between these halo mass distributions.  For example, the central
halo masses of active galaxies are larger by $\sim$0.15 dex compared
to inactive galaxies in the lowest stellar mass bin
($\log_{10}(M_*)\sim10.6$). Broadly speaking, however, the halo mass
distributions of active galaxies and inactive galaxies are remarkably
similar in this SAM. There is however one key difference between the
two samples. The AGN satellite fraction ($\sim$3\%) is an order of
magnitude lower than the satellite fraction for all galaxies in the SAM (note that this
difference is not obvious from Figure \ref{galacticus_mhalo_dis} which
shows probability density functions).

\begin{figure}
\begin{center}
\includegraphics[width=8cm]{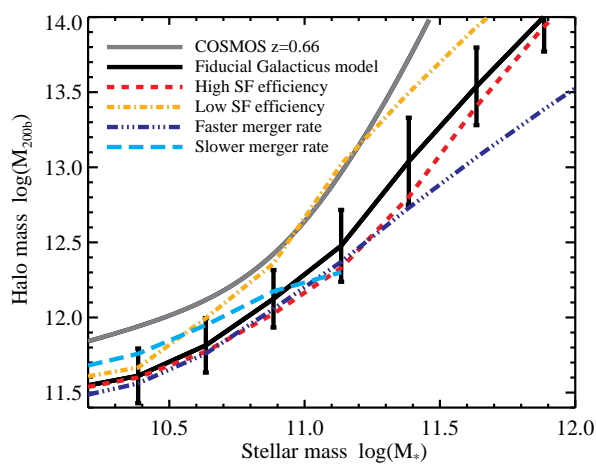}
\caption{Variations of the Galacticus SHMR with the star-formation
  efficiency and galaxy merger timescales. Errors on the SHMR indicate
  the 1-$\sigma$ scatter in the model relation at fixed stellar mass
  and are shown only on the fiducial model for visual purposes. In
  this SAM, modifying the accretion efficiency onto black holes in
  radio mode only has a minor effect on the SHMR and so these
  parameter variations are not displayed.}
\label{galacticus_shmr}
\end{center}
\end{figure}

\begin{figure*}
\begin{center}
\includegraphics[width=15cm]{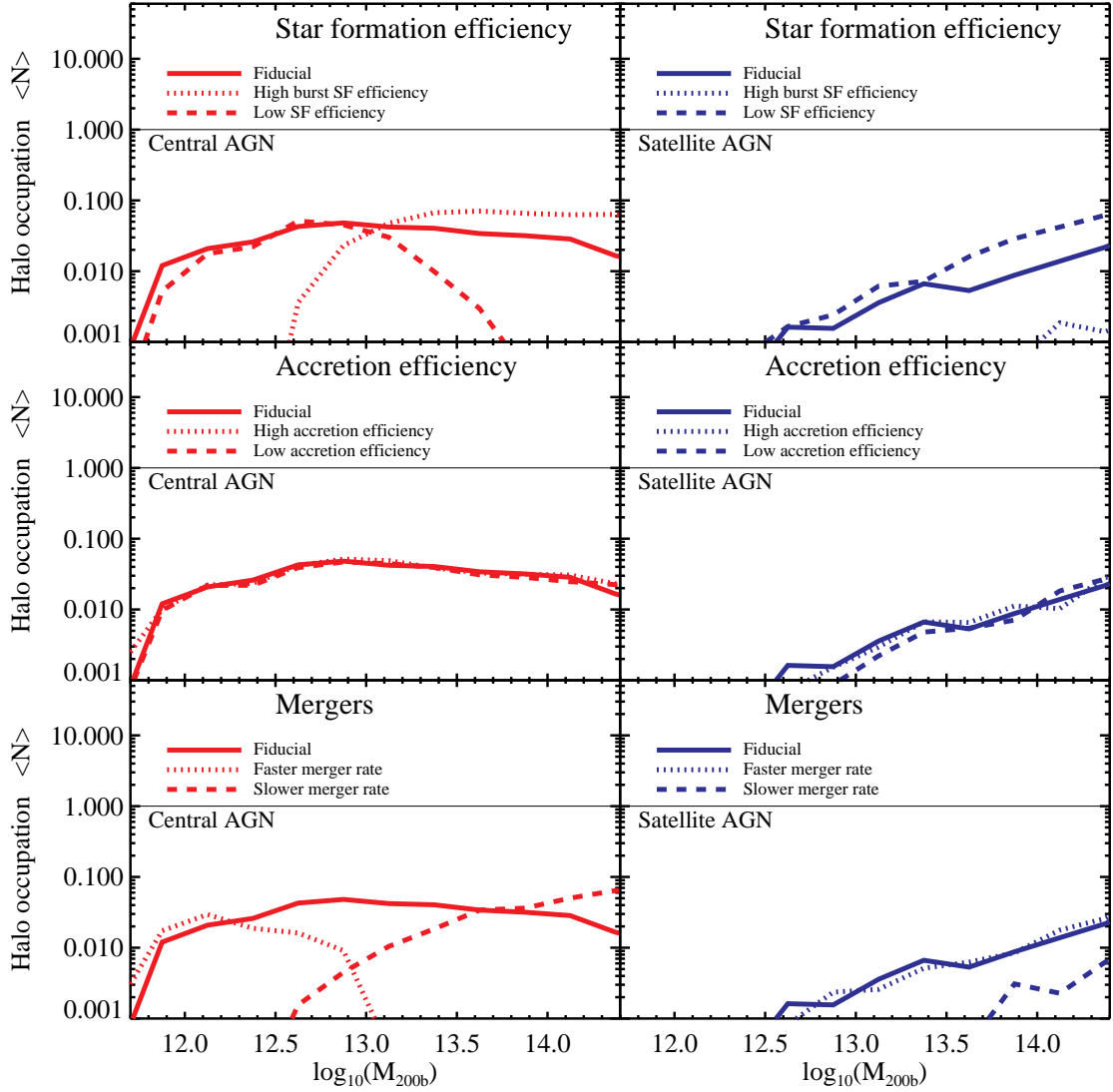}
\caption{Impact of a few key parameters on the HODs of AGN selected in
  the Galacticus SAM that have been roughly selected to mimic our
  COSMOS sample. Note that we do not necessarily expect the Galacticus
  HODs to match the ones derived in this paper because our base-line
  Galacticus model does not reproduce the COSMOS SHMR. Our goal in
  this figure is to compare qualitative variations in the occupation
  functions.}
\label{galacticus_hod}
\end{center}
\end{figure*}

The similarity between the halo mass distributions in Figure \ref{galacticus_mhalo_dis} supports our approach of using a fiducial
SHMR as a starting point to model this AGN population (at least when compared to this particular SAM). The low satellite fraction
of the active galaxy sample suggests, however, that in step three of our methodology (see section \ref{method}), $f_{\rm sat}$
should be left as a free parameter. The fact that this SAM predicts a much lower satellite fraction for active galaxies compared
to inactive galaxies also raises the possibility of an interesting tension between this SAM and our weak lensing results. However,
there are important differences between our data and this SAM that preclude a direct comparison. Above all, the SAM needs to
better match other global properties of the galaxy distribution given the expectation that the same physical processes that
regulate galaxy growth may also affect AGN activity. To first order, this requires matching the galaxy mass function and its
evolution with redshift --- a non-trivial task (recent progress on this topic is described by
\citealt[][]{Vogelsberger:2013,Benson:2014}).  In addition, the AGN weak lensing signal is more sensitive to high satellite
fractions than to low satellite fractions (Figure \ref{agnmocks}). An interesting direction for future work would be to use a
joint analysis of lensing and clustering to pin down $f_{\rm sat}$ with greater accuracy.

\subsection{Insights from Galacticus SAM: Physical Parameters that Regulate the AGN HOD}\label{sam2}

In Section \ref{hod} we derived the occupation functions for this AGN sample (Figure \ref{agn_hod}). However, HOD functions are
simply a stepping stone towards the broader goal of improving observational constraints on mechanisms that fuel AGN activity.
While a full discussion is beyond the scope of this paper (for a related discussion using hydrodynamic simulations see
\citealt{Chatterjee:2012}), in this section we provide a qualitative assessment of how AGN HODs relate to theoretical models of AGN
activity.

Figure \ref{galacticus_shmr} plots Galacticus SHMRs for X-ray AGN at $z\sim0.6$ with the relation derived from the COSMOS data
overlaid.  Given that the global SHMR in Galacticus does not match the data, we do not attempt a direct comparison but simply show
how physically-informative parameters of the semi-analytic model might be constrained by the AGN HODs. We vary the following key
parameters related to AGN activity in Galacticus and show their relative impact:

{\it 2) Star formation efficiency in bursts.} Models with efficiencies
$10$ and $1/10$ times the fiducial efficiency are considered. More
efficient star formation in bursts tends to reduce the net accretion
of gas onto black holes (since gas becomes more quickly locked up in
stars or ejected from the galaxy in winds).

{\it 3) Accretion efficiency onto black holes from the hot
  atmosphere.} Models with accretion from the hot atmosphere (which
drives the ``radio mode'' AGN) at rates $10$ and $1/10$ times the
fiducial rate are considered. The radio mode accretion rate controls
the efficacy of AGN feedback in the Galacticus model. We find however,
that modifying the accretion efficiency onto black holes in radio mode
does not affect the SHMR because the strength of radio model feedback
in this SAM is well above that required to completely shut down
cooling in high-mass halos. As such, reducing the accretion rate (and,
therefore, the feedback power) by a large factor still leaves enough
feedback power to shut down cooling, and increasing feedback power
makes no real difference (since once cooling is shut down, more
feedback can have no additional effect).

{\it 4) Galaxy merger timescales.} Timescales for galaxy-galaxy
mergers (driven by dynamical friction and with timescales computed
using the fitting formula of \citealt{Jiang:2008}) are varied
by a factor $10$ above and below the timescales in the fiducial
model. Rapid merging of galaxies leads to more rapid build up of black
hole masses (both by direct merging of black holes and by driving gas
into the spheroid where it may be accreted by the central black hole).

Figure \ref{galacticus_hod} displays the effects of varying these four
parameters on the Galacticus AGN HOD. One immediate point of interest
here is that the general {\em shapes} of the HODs from Galacticus
match those found in this paper quite well. Most of the HODs displayed
in Figure \ref{galacticus_hod} would be relatively well described by
Equations \ref{ncen} and \ref{nsat} for $N_{\rm cen}$ and $N_{\rm
  sat}$. However, although the shape of $N_{\rm sat}$ is well
described by a power-law, the overall amplitude of $N_{\rm sat}$ is
quite low. As discussed already in the previous section, this is a
manifestation of the fact that the satellite fraction for AGN is lower than
for galaxies in this SAM.

Figure \ref{galacticus_hod} shows that radio-mode accretion efficiency
has almost no effect on the AGN occupation functions. This is because
in this SAM, radio-mode accretion mainly dominates for AGN with lower
luminosities. On the other hand, star formation efficiency and mergers
have a large impact on the characteristic halo mass scales of the
central occupation functions as well as on the amplitude of the
satellite occupation function. Although we have not explored this
aspect in great detail, it is possible that a higher star formation
efficiency reduces the HOD in lower mass halos because gas is
efficiently used up by star formation instead of accreting onto the
black holes. A low star formation efficiency in bursts may result in
an enhanced satellite contribution because there is now more gas left
in satellite spheroids to accrete onto their black holes. Finally, a
low merger rate for galaxies may reduce the central galaxy HOD in low
mass halos due to the lack of major mergers which drive gas onto the
back holes.

However, as can be seen from Figure \ref{galacticus_shmr}, as we vary
the star formation efficiency in bursts and the Galaxy merger
timescales, the SHMR also varies. As a result, it is difficult to know
how much of the change in the model AGN HODs is due to changing the
nature of BH growth and activity and how much is due to simply
changing the SHMR. In practice, we would need to only explore models
with viable SHMRs to ascertain how these physical processes directly
affect the growth and fueling of black holes. This points to an
interesting direction for future research. The solution to this
problem will be to first calibrate the Galacticus model to accurately
match the measured SHMR. Using MCMC techniques as described in
\cite{Benson:2014} would allow us to survey the entire model parameter
space and locate those regions which adequately match the measured
SHMR. Sampling model parameters from these regions of parameter space
would then allow us to explore how the AGN HOD depends on model
parameters once the SHMR is fixed.


\section{Summary and Conclusion}\label{conclusions}
  
In this paper, we have developed a new framework for studying how black hole fueling may be tied to host dark matter halos by tying
measurements of AGN host stellar masses to prior knowledge about the SHMR. In contrast with previous work, which only
considered a single effective halo mass scale, the technique presented here can be used to infer the full halo mass distribution
for AGN samples.

HOD modeling of AGN populations is fundamentally limited by model degeneracies driven by the fact that AGN may live a wide range
of halo masses with an occupation function whose general shape and normalization are poorly known \citep[\eg][]{Shen:2013}. Faced
with this difficulty, we propose that instead of trying to constrain a full HOD model from AGN samples, we can ask a more simple,
but no less fundamental question: how do AGN samples differ from inactive galaxies of equivalent stellar mass? This can be
achieved through a rigorous comparison of the clustering, lensing, and cross-correlation signals of AGN hosts to the fiducial
stellar-to-halo mass relation (SHMR) derived for all galaxies, irrespective of nuclear activity.

The key advantage of this approach is that by using large samples of galaxies that are complete in terms of stellar mass, the SHMR
can be built with much higher accuracy than by using any statistic measured from AGN samples alone. Statistics measured from AGN
are only used to constrain {\em deviations} from the fiducial model.

We have applied this technique to a sample of moderate luminosity ($\langle \log(L_X) \rangle = 42.7$) obscured X-ray AGN at $z<1$
from the COSMOS field.  Despite the small sample size (several hundred AGN) we demonstrate that our method can be used to
constrain AGN halo occupation statistics. For the first time, we measure the galaxy-galaxy lensing signal of X-ray selected
obscured AGN. We find excellent agreement between the AGN lensing signal and the prediction based on our fiducial SHMR. 
There is no evidence from our analysis to suggest that AGN populate dark matter halos in a different manner compared to galaxies with
the same $M_*$, regardless of nuclear activity. We discuss how similar tests in future work could equally well be
performed for the AGN auto-correlation function, or for cross-correlations between AGN and mass-limited galaxy samples. 

In contrast with previous work which typically only provides a single
effective halo mass scale, the technique presented here can be used to
infer the full halo mass distribution for AGN samples. Contrary to
conventional wisdom, our method suggests that most X-ray AGN do not
live in medium sized groups with $M_{\rm h} \geq 10^{13}
M_{\odot}$. Instead, 50\% of the AGN in our sample lives in halos less
massive than $\log_{10}(M_{\rm 200b})\sim12.5$ and hence in relatively
low-mass dark matter halos. Only $\sim$60\% of AGN satellites are
contained in halos with $\log_{10}(M_{\rm 200b})>13$. We stress that
these values are specific to our particular AGN sample selection and
that the lower halo mass limit described here is primarily set by our
choice to select an AGN sample with host masses above
$\log_{10}(M_*)>10.5$.  Our work is consistent with moderate
luminosity X-ray AGN occupying a wide range of halos masses. A similar
picture is supported for luminous QSO samples by
\citet[][]{Conroy:2013}. Taken together, these two results suggest
that both QSOs and moderate luminosity X-ray AGN may occupy halos in a
relatively ``normal'' way, calling into question previous claims for
an environmental signature of distinct fueling modes for QSOs compared
to moderate luminosity X-ray AGN.

We compare our results with previous halo mass estimates inferred from X-ray clustering. We globally find that our predicted
effective halo mass (measured in a consistent fashion as with clustering studies) is lower than previous work. However, we also
caution that sample selection effects may be non negligible when performing such comparisons and need to be considered
carefully. Studies of X-ray AGN clustering are typically limited to bright hosts simply due to the fact that such measurements
require spectroscopic redshifts. As a result, samples from previous work are typically limited to hosts
with I$_{AB}<$21.5-23. In detail, there are important variations in the selection functions applied to AGN samples between
different studies. Differences in the dark matter halo distributions between various groups are in fact expected---hosts
with brighter luminosities on average live in larger dark matter halos. A fixed I-band cut will also probe different host
stellar masses at different redshifts. We stress that these (sometimes implicit) cuts on host properties need to be accounted
before meaningful comparisons can be made.

We derive the halo occupation functions for our sample and show that they are well described by the same functional form for
galaxies but with an overall amplitude normalization of $f_A\sim0.028$. At group scales, the satellite occupation is well
described by a power-law with a slope of $\alpha=1$.

Finally, we investigate some simple models from the Galacticus SAM and find broadly consistent shapes for AGN HODs. However, in
contrast with our lensing results, the SAMs predict an AGN satellite fraction that is an order of magnitude lower compared to the
overall galaxy population. This suggests a tension worth investigating in future work using higher S/N weak lensing and clustering
measurements for AGN host galaxies.

\section*{Acknowledgements}

We thank Phil Hopkins, Surhud More, and John Silverman for insightful
discussions while preparing this paper. We thank Ed Turner for
valuable discussions related to statistical methods. We are grateful
to Ian Harnett for editing this manuscript. We thank Nikos Fanidakis
for clarifications regarding halo mass values in
\citet[][]{Fanidakis:2013}. This work was supported by World Premier
International Research Center Initiative (WPI Initiative), MEXT,
Japan. AK is supported by the National Science Foundation Graduate
Research Fellowship, Grant No. DGE-1148900. FC acknowledges financial
support by the NASA grant AR1-12012X. RM is supported by a Royal
Society University Research Fellowship. JR was supported by JPL, which
is run by Caltech under a contract for NASA. ALC acknowledge support
from NSF CAREER award AST-1055081.

\appendix
\section{The Galacticus SAM}

In Galacticus, black holes are assumed to accrete from both the
interstellar medium in the spheroid of their host galaxy and the hot
atmosphere of gas surrounding the host galaxy at rates governed by
Bondi-Hoyle accretion \citep{Edgar:2004} with a multiplicative
pre-factor designed to take into account the fact that the model does
not resolve the relevant length scales for accretion
\citep{Booth:2009}. The nature of the accretion disk surrounding each
black hole is determined by the accretion rate onto the black hole. At
accretion rates below 1\% or above 30\% of the Eddington accretion
rate the accretion disk is modeled as a radiatively inefficient,
geometrically thick ADAF \cite{Narayan:1994}, otherwise a radiatively
efficient, geometrically thin \cite{Shakura:1973} solution is
used. The evolution of black hole spin is also tracked, using the
method of \cite{Benson:2009} to account for spin-up by accretion and
spin-down by jet production. During galaxy mergers, black holes are
assumed to merge instantaneously. The resulting merged black hole has
a mass equal to the sum of the masses of its progenitors and a spin
computed using the method described by \cite{Rezzolla:2008} assuming
that the progenitor black holes have randomly aligned spin vectors. In
comparison with the black hole evolution model of
\cite{Fanidakis:2011}, our model ignores the details of misaligned
accretion disks--black hole spins, but employs a more detailed model
of accretion\footnote{In \protect\cite{Fanidakis:2011} a fixed
  fraction of the available gas mass is funneled into black holes
  during each galaxy merger or disk instability event.}. In other
respects, our model and that of \cite{Fanidakis:2011} are comparable
in terms of the physics included and level of detail in the modeling.

Each galaxy therefore contains a supermassive black hole with known
mass, spin, and accretion rate. With the default parameters of our
model the correlation between black hole mass and spheroid stellar
mass \citep{Haring:2004} is approximately reproduced. The
bolometric luminosity is computed from the black hole rest mass
accretion rate and radiative efficiency (assumed to be $\epsilon_{\rm
  rad}=1-E_{\rm ISCO}(j)$ for a black hole of spin $j$ accreting via a
thin accretion disk, where $E_{\rm ISCO}$ is the specific energy of
material at the innermost stable circular orbit of the black hole,
while for a black hole accreting from a radiately inefficient thick
accretion flow the radiative efficiency is $0.01 \lambda/\lambda_{\rm
  thin}$ where $\lambda$ is accretion rate in units of the Eddington
rate, and $\lambda_{\rm thin}$ is the minimum such accretion rate at
which a thin disk occurs). An SED for an AGN of this bolometric
luminosity is then computed using the model of
\cite{Hopkins:2007a}. From this SED, a broad band luminosity is
computed in each X-ray band assuming a spectrum of the form $f_\nu
\propto \nu^\alpha$ with $\alpha=-0.4$ as in the observational
analysis.

Our COSMOS AGN sample is expected to be roughly obscured by a mean
column density of $N_H \sim10^{22}$ cm$^{-2}$ with values extending
out to $N_H \sim10^{23}$ cm$^{-2}$ (see Figure 2 in
\citealt[][]{Lusso:2011}). \citet{Lusso:2011} find a mean value of
$N_H \sim10^{22}$ cm$^{-2}$ for a similarly selected sample of Type-2
AGN from the XMM-COSMOS sample. In this Galacticus SAM, the X-ray
luminosity, $L_X$, is attenuated from a fixed overall column density
(mimicking a torus + ISM) of $N_H=$10$^{22}$ cm$^{-2}$ assuming solar
metallicity. The photoelectric absorption cross-section per hydrogen
is computed as a function of photon energy using
\cite{Wilms:2000}. Multiplying by the hydrogen column density gives
the net absorption as a function of energy to the AGN.


\bibliographystyle{mn2e}

\label{lastpage}


\end{document}